# Triple Scoring Using a Hybrid Fact Validation Approach

## The Catsear Triple Scorer at WSDM Cup 2017

Edgard Marx[*]
AKSW, University of Leipzig
HTWK, Leipzig University of Applied Sciences
marx@informatik.uni-leipzig.de

Tommaso Soru[*]
AKSW, University of Leipzig
tsoru@informatik.uni-leipzig.de

André Valdestilhas
AKSW, University of Leipzig
valdestilhas@informatik.uni-leipzig.de

## ABSTRACT

With the continuous increase of data daily published in knowledge bases across the Web, one of the main issues is regarding information relevance. In most knowledge bases, a triple (i.e., a statement composed by subject, predicate, and object) can be only true or false. However, triples can be assigned a score to have information sorted by relevance. In this work, we describe the participation of the **Catsear** team in the Triple Scoring Challenge at the WSDM Cup 2017. The Catsear approach scores triples by combining the answers coming from three different sources using a linear regression classifier. We show how our approach achieved an Accuracy2 value of $79.58\%$ and the overall 4th place.

## 1. INTRODUCTION

The technological advances in the information era led to a tremendous volume of structured information.This data is continuously generated by humans, as well as by different types of smart agents. For instance, as of today more than 10,000 *Resource Description Framework (RDF)*[1] knowledge bases are publicly available.[2] Portals as WorldWideWebSize[3] report more than 50 billion of Web pages on the Web.

Especially within very large knowledge bases, it is hard to find relevant information. If on the one hand the total lack of information leaves users uninformed, on the other hand, the use of all available unsorted information can lead to misinformation. However, relevance can be perceived differently depending on the individual and the context. In this work, we present a hybrid method to score triples based on multiple information sources. Inspired by Fact Validation approaches, we learn a trustworthiness level for each source and combine their outcomes to yield a final relevance score for a triple. We describe how our approach *Catsear* achieved the *4th place* in the Triple Scoring task at the WSDM Cup co-located with WSDM 2017 [5].

An extensive list of related works can be found in the WSDM Cup overview paper [3]. Triple Scoring is often related to the field of Triple Classification. Triple Classification approaches provide a confidence score for a triple within a given set of knowledge bases [6, 13]. This specific work is also related to other approaches for fact validation. Fact Validation instead relies on external sources, processing structured and unstructured data. For instance, DeFacto [4] is an approach able to validate facts by finding trustworthy sources for them on the web. DeFacto collects and combines evidence from web pages written in several languages, also including support for facts with a temporal scope.

The remainder of this paper is organized as follows. We briefly introduce the challenge task in Section 2. In Section 3, we describe *Catsear* in detail, presenting its components from Section 3.1 to Section 3.4 and the implementation in Section 3.5. We then show some of the results we obtained in Section 4. The rest of the evaluation containing the part which was eligible for the challenge is shown in [3]. In Section 5, we conclude.

## 2. TASK

The challenge featured two datasets, containing only predicates about nationalities and professions, respectively. Each dataset is divided into training and test sets. The training and test set sizes for professions are of $515$ and $343,329$, while for nationalities are of $162$ and $301,590$, respectively. Participants were expected to return an integer value in $[0, 7]$ for each triple in the test sets, which were not directly accessible. Further details on the available data can be found in [2].

## 3. APPROACH

We firstly analyzed our training sets. In many examples, we were able to find scores which appeared controversial and open to interpretation. For instance, in the nationality training set[4], triples with subject *Frederick Loewe* feature the following objects and receive the following scores: (*United States of America*, $7$), (*Austria*, $5$), and (*Germany*, $3$). However, according to Wikipedia[5], *Frederick Loewe* was born in *Berlin* to *Austrian* parents. Although it is possible to see that his career has developed in the *United States of America*, there are more facts supporting *Austria* as his possible nationality than otherwise. Given these premises, we decided to develop an approach based on multiple sources to allow for the heterogeneity of interpretations.

The Catsear architecture can be seen in Section 3. Our triple scorer is based on four main modules plus two libraries distributed in three layers. The bottom layer represents the libraries used to handle data, i.e. *KBox* [9] and the *Word2Vec* [10] library for word embeddings. The middle layer involves three modules, named *\*Path* (read *Starpath*), *Graph Cross*, and *Skip Gram*. A *Super Classifier* occupies the top of the stack. Each of the middle-layer modules runs independently and provides the upper module with a prediction of the triple score. The final score is then outputted by the Super Classifier.

---

[*]These authors contributed equally.
[1]http://www.w3.org/RDF
[2]http://lodstats.aksw.org/
[3]http://www.worldwidewebsize.com/
[4]http://broccoli.cs.uni-freiburg.de/wsdm-cup-2017/nationality.train
[5]https://en.wikipedia.org/wiki/Frederick_Loewe

| Super Classifier | | |
|---|---|---|
| *Path | Graph Cross | Skip Gram |
| KBox | | Word2Vec |

Table 1: The Catsear architecture.

## 3.1 The *Path module

*Path is part of previous research published in [8] and was originally developed for Question Answering on knowledge graphs and graph disambiguation. In this challenge, the approach was used to validate the given predicates (i.e., *profession* and *nationality*) using the DBpedia [1] and Yago [7] knowledge graphs. Before executing the *Path algorithm, we locally installed DBpedia and Yago Labels using *KBox*, a solution to provide ready-to-use knowledge graphs shifting the query executions to the end users [9].[6] KBox allows users to collect datasets from a global distributed repository. Datasets are then ready to be queried by any application.

We locate the entity by adding the label to the DBpedia namespace (e.g. `http://dbpedia.org/resource/`). Thereafter, we retrieve the entity's Semantic Connected Component (SCC) (see Definition 3) and check for the existing object in one of the DBpedia predicates $P'$ seen in Table 2. In short, the SCC of an entity $e$ are all triples $(e, p, o)$ and their respective labels, after applying the forward reasoning. This allows, for example, to know that an entity typed as `:Politician` is also a `:Person`, even when the relation is not declared.[7]

In order to understand how the *Path approach works, we introduce the following definitions. From this point on, we consider an RDF graph as a labelled directed graph $G = (V, E)$ where for any triple $(s, p, o) \in G$ we have $s, o \in V$ and $p(s, o) \in E$.

DEFINITION 1 (NATURAL LANGUAGE QUERY). *An NL query $q \in \Sigma^*$ is a user given keyword string expressing a factual information needed.*

DEFINITION 2 (TERM NETWORK). *A Term Network is a graph whose vertices are labeled with terms.*

The *Term Network* is extracted from a structure called *Semantic Connected Component* (SCC).

DEFINITION 3 (SEMANTIC CONNECTED COMPONENT). *The Semantic Connected Component (SCC) of an entity given by a subject $s$ in an RDF graph $G$ under a consequence relation $\models$ is defined as $SCC_{G,\models}(s) := \{(s, p, o) \mid G \models \{(s, p, o)\}\} \cup \{(p, rdfs{:}label, l) \in G\} \cup \{(o, rdfs{:}label, l) \in G\}\}$. From this point on, we use the shorter notation $SCC(s)$.*

Then, a model called *Semantic Weight Model* (SWM) is applied to the tokens extracted from the entity's Term Network.

DEFINITION 4 (TOKEN). *A token $t \in T$ is the result of a tokenizer function $T : \Sigma^* \to \Sigma^{**}$, which converts a string into a set of tokens.*

DEFINITION 5 (SEMANTIC WEIGHT MODEL, SWM). *Each token $t$ in $T(q)$ is firstly mapped to the paths of the SCC $S$. The set of matched tokens from a path $\gamma$ is returned by the function* $TP(\gamma, q)$. *A path match of an SCC $S$ is evaluated by the function* $\text{MTP}(\gamma, q, S)$ *using a path weighting function* $w : E^+ \to R$.

---
[6]KBox is available at http://kbox.tech.
[7]This is known as RDF entailment; https://www.w3.org/TR/sparql11-entailment/.

$$TP(\gamma, q) := \{t \in T(LP(\gamma)) \mid \exists t' \in T(q) : \delta(t, t') < \theta\} \quad (1)$$

$$\text{MTP}(\gamma, q, S) := \{t \in TP(\gamma, q) \mid \forall \gamma' \in E(S)^+ : \quad (2)$$
$$w(\gamma)|TP(\gamma, q)| \geq w(\gamma')|TP(\gamma', q)|\} \quad (3)$$

*The final score of an SCC $S$ is a sum of its $n$ path-scores and is measured by the function* $\text{score}(S)$ *as follows:*

$$\text{score}(S, q) = \sum_{\gamma \in E(S)^+} \begin{cases} w(\gamma)|TP(\gamma, q)| & \text{if } \text{MTP}(\gamma, q, S) \neq \emptyset, \\ 0 & \text{otherwise.} \end{cases} \quad (4)$$

*In case there are terms matching multiple paths and the paths have an equal number of matched terms and equal score, only one of the path scores is added to the SCC score.*

For scoring a given predicate, we design three evidence levels: (1) full, the predicate was found; (2) partial, the predicate was partially found, and; (3) none, the predicate could not be found. We scored a full, partial, and non-evidence respectively with 5, 3, and 2. The evidence for each of the input predicates $p'_W \in P'_W$ is given by its respective maximal DBpedia predicates $P'$ score, considering $o'_W \in O'_W$ as input object, the score of a tuple $(s, p'_W, o'_W)$ is given as follows:

$$SCC_W(s, p'_W) = \{(s, p', o) : p' \in dbpedia(p'_w)\}$$

$$score_{*Path} = score(SCC_W(s, p'_W), o'_W)$$

$$score_1(s, p'_W, o'_W) = \begin{cases} 5 & \text{if } score_{*Path} \geq 1, \\ 3 & \text{if } 1 > score_{*Path} > 0, \\ 2 & \text{otherwise.} \end{cases} \quad (5)$$

This value, along with the outputs of the other middle-layer modules, are then forwarded to the Super Classifier.

## 3.2 The Graph Cross module

The aim of the Graph Cross module is to estimate the triple scores using ProBase [14], a taxonomy of concepts for short text understanding, which is integrated at the time of writing into a project dubbed Microsoft Concept Graph.[8] As the Concept Graph is available only in TSV files, we firstly converted the knowledge base into RDF. The knowledge base only contains type relationships weighted by an integer number. We thus normalized and rounded the weights into $\{1, ..., 7\}$ and created an RDF property for each new weight (e.g., `/property/type/7`). We call this weighting function $ms : p \to \{1, ..., 7\}$.

Thereafter, we installed it on the local machine using KBox. Before sticking to KBox, we tried other solutions to handle the 1.4 GB knowledge base. Loading and indexing the information in-memory showed to saturate the available memory of our machine, whilst using a DBMS-based approach showed to be extremely slow. However, the KBox-based method performed better in terms of memory consumption and runtime.

Besides the knowledge base, we make use of a list of demonyms. A demonym is a proper noun used to denote the natives or inhabitants of a particular country, state, or city. For example, German is a demonym for people from Germany and Brazilian is a demonym for people from Brazil. We gathered this information from DBpedia 2016-04 using the following SPARQL query.

---
[8]https://concept.research.microsoft.com

| WSDM Predicate | DBpedia Properties |
|---|---|
| Nationality | dbp:birthplace dbo:birthplace dbp:cizenship |
| Profession | rdf:type dbp:profession dbo:profession dbo:occupation |

**Table 2: DBpedia properties used by *Path in each of the WSDM predicates. `dbo:` and `dbp:` are short forms for the DBpedia ontology and property namespaces.**

```
SELECT str(?c) str(?d) WHERE {
    ?x a dbo:Country .
    ?x dbo:demonym ?d .
    ?x rdfs:label ?c .
    FILTER(lang(?c) = "en")
} ORDER BY ?c
```

Afterwards, we manually cleaned the data. The list of countries and demonyms has been made available in the Catsear open-source proceedings repository.

The Graph Cross section consists of three steps:

1. Load all records from the training set organized as a hash structure of key and value, where the key represents the subject and the value is a sub-hash where the key is the predicate and the value is the score: $hashWSDM = \langle subject \langle predicate, score \rangle \rangle$.

2. Perform two predictions for the triple score: one using the objects of the given triple; one (applicable to predicate *nationality* only) using the demonyms.

3. Generate a file with the new score predictions ready for being processed by the Super Classifier.

The first prediction value of this module is thus defined as:

$$score_2(s,p,o) = \begin{cases} ms(p') & \text{if } (s,p',o) \in G \\ 0 & \text{otherwise.} \end{cases} \quad (6)$$

where $G$ is the RDF version of the Microsoft Concept Graph. Note that there cannot exist more than one triple with the same subject and object.

---

**Algorithm 1** The Graph Cross algorithm.

**Input**: $hashWSDM = \langle name\langle predicate, score \rangle \rangle$, $hashModel = \langle name\langle predicate, score \rangle \rangle$.
**Output**: $hashPredict = \langle name\langle predicate, score \rangle \rangle$.

1: **procedure** GRAPHCROSS()
2:    $demonyms$ = loadDemonyms($fileDemonyms$)
3:    **for all** $name \in hashWSDM$ **do**
4:       $kboxProp$ = kbox.getObjects($name$)
5:       **for all** $\langle attribute, score \rangle \in name$ **do**
6:          **for all** $p \in kboxProp$ **do**
7:             **if** $\exists p \in demonyms$ **then**
8:                push onto $hashPredict$ name, predicate, $(score \in p) + 1$
9:             **end if**
10:          **end for**
11:       **end for**
12:    **end for**
13:    **return** $hashPredict$
14: **end procedure**

---

The second prediction value is defined as Equation (7) shows.

$$score_3(s,p,o) = \begin{cases} a \cdot d & \text{if } d < \frac{7}{a} \\ 7 & \text{otherwise.} \end{cases} \quad (7)$$

where $0 < a \leq \frac{7}{d}$ is a parameter and $d$ is the number of occurrences of the demonym found in the Microsoft Concept Graph. For instance, finding `German scientist` for Albert Einstein would add up 1 to $d$ for Germany. We found the optimal parameter value of $a = 3$ by maximizing the Accuracy2 for this module on the training set.

### 3.3 The Skip Gram module

#### 3.3.1 Learning phase

As the module name suggests, we chose the Skip-Gram neural-network model to represent words in a vector space. The Triple Score challenge came with a dataset of 33,159,353 sentences from Wikipedia with annotations of the persons who figured as subjects of the input triples. We processed this text file applying a regular expression to transform all annotations such that only the entity name is left (e.g., from [Núria_Espert|Núria Espert] to Núria_Espert). This way, the Word2Vec model treats an entity as a word, assigning a vector to it. We trained the Word2Vec model on the text corpus, selecting a vector size of 300, a window span of 10, and a minimum number of occurrences for a word of 3.

#### 3.3.2 Score prediction

For each input triple, we extracted the subject (i.e., person name) and replaced all spaces with an underscore (_). Performing a lookup in the Word2Vec dictionary, it is possible to check the existence of a vector associated with it. In case the vector does not exist, the module outputs a value of 7. Otherwise, we compute the cosine similarity $\sigma$ between the subject and every possible object for the current predicate. Also these values were available in the task. The final score for the Skip Gram module is finally given by:

$$score_4(s,p,o) = 2 + 5 \frac{\sigma(s,o) - m}{M - m} \quad (8)$$

where $m$ and $M$ are the minimum and maximum similarity value found within all possible objects. As can be seen, the score was normalized into the interval $[2, 7]$.

For predicate *nationality*, besides the canonical score obtained with the simple country name, we also used the similarities between the subject and the demonyms of the given country. In case of multiple demonyms (e.g., "Polish" and "Pole" for Poland), we kept the maximum similarity value. We call this output $score_5$; it assumes a value of 0 for predicate *profession*.

### 3.4 The Super Classifier module

#### 3.4.1 Learning phase

We collected all five output values from the three modules for each instance in the training set. All middle-layer modules use an unsupervised-learning method, meaning they do not make use of the training labels. On the other hand, the Super Classifier module is composed by a Linear Regression classifier with Ridge Regularization [11] and a binary classifier on top. We thus considered the output values as features of our training set and learned the regression weights and the final threshold using 10-fold cross-validation.

Data from both predicates was merged to form one only training set to avoid overfitting and increase model robustness. The weights $w = [0.5245, 0.4532, 0.3513, 0.3824, 0.3824]$ and $b = -0.5606$ were obtained by the Linear Regression using Ridge Regularization parameter $R = 10^{-8}$. The values in $w$ correspond to the trustworthiness scores of the respective sources. On 677 instances, the correlation coefficient between predicted and real scores was $\rho = 0.4758$.

At the very last step, we classified the instances into two classes, in order to trig the output values only to 2 and 5. This step showed to bring benefit to the final Accuracy2 score. We trained our binary classifier to find a threshold $\tau \in [0, 7]$ which maximizes the Accuracy2 scores for both predicates. The value we found was $\tau = 3.5$.

### 3.4.2 Classification phase

In classification phase, the formula we used to estimate the score for a triple $(s, p, o)$ can be expressed as:

$$score_{LR}(s,p,o) = \sum_{i=1}^{5} w_i score_i(s,p,o) + b \qquad (9)$$

Subsequently, we applied the last classifier as:

$$score(s,p,o) = \begin{cases} 5 & \text{if } score_{LR}(s,p,o) \geq \tau, \\ 2 & \text{otherwise.} \end{cases} \qquad (10)$$

assigning a value of 5 or 2 to each triple in the test set.

## 3.5 Implementation

*Catsear* is a hybrid triple scorer with no common programming language among all modules. The *Path and Graph Cross modules were implemented in Java 8. The Skip Gram and Super Classifier modules and the scripts for the learning phases were implemented in Python 2.7. Finally, a Bash script was created to orchestrate the calls to the modules; it allows to run *Catsear* on one or more datasets. All code and links to the used material have been published online in the official open-source proceedings repository.[9] Experiments were carried out on a single-core virtual machine with 4 GB RAM running Ubuntu 16.04, made available through the TIRA interface for reproducible research [12].

## 4. EVALUATION RESULTS

In Section 3.4.1, we presented how we trained the Linear Regression classifier on the set of features coming from the respective modules. In order to choose our classifier, we ran an evaluation for N different classifier using 10-fold cross-validation. Linear Regression showed to achieve a higher *Accuracy2*[10] value than the J48 tree learner, Support Vector Machines, and Multilayer Perceptrons.

We evaluated the *Catsear* approach in the Triple Scoring Challenge at WSDM Cup 2017 [2, 3]. The tasks consisted in scoring triples from datasets containing persons' nationalities and professions. The participants were free to use data from any source as well as any amount of computation. However, the competitors should not make use of users' judgments.

The training phase consisted of evaluating 515 triples (pertaining to 134 persons) from professions and 62 triples (pertaining to 77 persons) from nationalities. Afterwards, the users were asked to set up their running approaches in a virtual machine using TIRA. After submitting the working systems, the users could select the dataset to evaluate. There were no limits on the number of experiments allowed for the participants. The final results were delivered after the competition was over. Table 3 shows the top seven teams and their resp. achieved position and Accuracy2 score. On the test set, the *Catsear* approach achieved the 4th position with an *Accuracy2* value of 79.58%, an Average Score Difference (*ASD*) value of 1.86 and a Kendall's *TAU* value of 0.41. While our method achieved a competitive value of *Accuracy2*, its *ASD* value ranked only 11th. This can be explained by the fact that the possible output scores for some modules were set distant from the ends of the score range (i.e., far from 0 and 7). Moreover, the application of a binary classifier on top contributed to an increase of the *ASD* value. On the training sets, averaged on both datasets, we recorded an *ASD* value of 1.82 without and 1.85 with the binary classifier.

---
[9] https://github.com/wsdm-cup-2017/catsear
[10] Accuracy on the score prediction with an error margin of 2, inclusive.

| # Position | Participant | Accuracy2 (%) |
|---|---|---|
| 1 | Bokchoy | 86.76 |
| 2 | Lettuce | 82.25 |
| 3 | Radicchio | 79.72 |
| **4** | **Catsear** | **79.58** |
| 5 | Samphire | 78.03 |
| 6 | Cress | 77.89 |
| 7 | Chickweed | 77.18 |

Table 3: Top-7 participants of the Triples Scoring challenge, WSDM Cup and their respective Accuracy2 values.

| Module | Nationality | Profession | Source |
|---|---|---|---|
| *Path | 72% | 55% | DBpedia (RDF) |
| Graph Cross | 15% | 43% | Wikipedia (Text) |
| Skip Gram | 73% | 68% | Microsoft CG (Triples) |
| **Super Classifier** | **88%** | **77%** | - |

Table 4: Accuracy2 (%) achieved by each module on the training set.

Table 4 shows the *Accuracy2* of each individual *Catsear* module, as well their combination using a *Super Classifier* with different datasets (i.e., Nationality and Profession), using the training data. Note that the enhancements using the country demonyms are not considered in the individual results. The Super Classifier values are obtained using 10-fold cross-validation.

## 5. CONCLUSION

In this paper, we presented a hybrid approach for triple scoring that combines results from three different sources. This approach makes use of the `*Path`, `Graph Cross` and `Skip Gram` modules to gather information from the sources and a `Super Classifier` module to learn the trustworthiness scores associated with them. We achieved the 4th position with an accuracy level of 79.58% at the Triple Score Challenge of the WSDM Cup 2017.

## 6. ACKNOWLEDGMENTS

This work was partly supported by a grant from the German Research Foundation (DFG) for the project *Professorial Career Patterns of the Early Modern History: Development of a scientific method for research on online available and distributed research databases of academic history* under the grant agreement No GL 225/9-1, by CNPq under the program *Ciências Sem Fronteiras* and by the *Instituto de Pesquisa e Desenvolvimento Albert Schirmer* (CNPJ 14.120.192/0001-84).